\documentclass[12pt,draftclsnofoot, onecolumn]{IEEEtran}

\usepackage{soul}
\usepackage{color}
\soulregister\cite7
\soulregister\ref7
\soulregister\eqref7
\soulregister\susubsection7
\soulregister\it7

\usepackage{amsmath,amssymb,amsthm}
\usepackage{graphicx}
\usepackage{makecell}
\usepackage{multirow}
\usepackage[hyphens]{url}
\usepackage{pdflscape}
\usepackage{afterpage}
\usepackage[utf8]{inputenc}
\usepackage[english]{babel}
\usepackage{amsthm}
\usepackage{amsmath}
\usepackage{enumerate}
\usepackage{wrapfig}
\usepackage{cite}
\usepackage{caption}
\usepackage{subcaption}

\newtheorem{theorem}{Theorem}
\newtheorem{corollary}{Corollary}

\theoremstyle{definition}
\newtheorem{definition}{Definition}
\newcommand\ddfrac[2]{{\displaystyle\frac{\displaystyle #1}{\displaystyle #2}}}

\begin{document}

\title{Separation of Control and Data Transmissions in 5G Networks may not be Beneficial}

\author{\IEEEauthorblockN{Zainab Zaidi\IEEEauthorrefmark{1}, Hazer Inaltekin\IEEEauthorrefmark{2} and Jamie Evans\IEEEauthorrefmark{3}\\}
\IEEEauthorblockA{\IEEEauthorrefmark{1}IEEE Senior Member, Email: zzaidi@ieee.org\\}
\IEEEauthorblockA{\IEEEauthorrefmark{2}School of Engineering,
Macquarie University\
Email: hazer.inaltekin@mq.edu.au\\}
\IEEEauthorblockA{\IEEEauthorrefmark{3}Department of Electrical and Electronics Engineering\\
University of Melbourne, Email: jse@unimelb.edu.au\\}\vspace{-10mm}
}

\maketitle

\begin{abstract}
The logical separation of control signaling from data transmission in a mobile cellular network has been shown to have significant energy saving potential compared with the legacy systems. As a result, there has been a lot of focus in recent years on development and realization of separation architectures. Our study, however, shows that the energy savings of separation architecture remain under $16-17\%$ when compared with legacy systems and this gain falls to a mere $7\%$ when both architectures are realized under a CloudRAN (CRAN) setting. Moreover, when we strategically place some small base-stations (SBSs) to cover the area in a densely deployed scenario and allow all other base-stations (BSs) to be used only on-demand, the system consumes much less energy than the separation architecture. While we expected that most equipment would be shut down during nights, our study shows that around $70\%$ of the small cells are required to be active to serve randomly distributed minimum data load, i.e., active mobile equipment. Contemporary mobile traffic is predominantly data which does not go to extremely low levels during nights. We discuss, in detail, the assumptions, their implications, and the effects of system parameter values on our conclusions.  
\end{abstract} 
\begin{IEEEkeywords}
Control-Data Separation Architecture (CDSA), dense deployment, CloudRAN, energy management
\end{IEEEkeywords}
\vspace{-3mm}
\section{Introduction}
Separation of control and data transmissions is suggested as the ultimate architectural design for cellular networks to achieve the real benefit of energy management \cite{Rittenhouse2012, Godor2012, Filippini2017}. In the traditional cellular systems, the network is required to transmit mandatory control signals to provide always-{\it on} connectivity even when no data is transmitted \cite{Godor2012}. When coverage is separated from capacity, through logical decoupling of data and control transmissions, the equipment responsible for data, and consuming most of the power, can only be used on-demand. Here, the signaling BSs provide always-{\it on} connectivity and will be designed for low rate services consuming a very small fraction of the power \cite{Zaidi2015,Filippini2017}. The separation architecture discussed in this paper, also called CDSA (Control-Data Separation Architecture), is about making radio access part of the network (RAN) more energy efficient \cite{Mohamed2016} and should not be confused with separation of control and user plane (CUPS) in EPC or core network in the context of 5G (https://www.3gpp.org/cups) to enable the implementation of Software Defined Networking for enhanced scalability. The state of the art of CDSA considers a macro cell as a control or signaling only cell and data transmissions are managed by SBSs \cite{Pat2019,Taufique2019}.

Earlier studies suggest that control-data decoupling may provide $85-90\%$ energy saving potential compared to the legacy systems (LTE) \cite{Godor2012,Zaidi2015} and remarkable improvement in energy efficiency \cite{Filippini2017,Mukherjee2013,Ternon2013,Wang2014,Kang2017}. There has been significant focus towards development and realization of this paradigm \cite{Mohamed2016,Taufique2019}. Our study, however, shows that the logical split of signaling and data, referred to as separation architecture in this paper, is not always the most energy efficient when compared with legacy or non-separation architectures under reasonable assumptions for operating conditions. By legacy architectures, we refer to the networks consisting of standard macro BSs and SBSs, e.g., macro eNodeB/gNB and small eNodeB/gNB, which are providing control signaling as well as data transmissions. We observed that legacy networks with an appropriate energy management scheme and a small number of strategically placed SBSs to provide coverage, can provide much better energy savings than separation architecture. The major cause for marginal savings is the requirement to activate around $70\%$ of the small cells in a separation architecture to serve the randomly distributed minimum load during early morning hours, where full load activates $100\%$ small cells. We looked into each of the earlier studies to find out the reasons for this huge discrepancy. Section \ref{rel-work} has our detailed comments. In short, we found that the major reasons for inflated saving estimates are the following.
\begin{enumerate}
\item Mobile traffic has become predominantly data. While voice traffic almost goes to $0\%$ during nighttime, data traffic remains over $10\%$. Considering the spatial distribution of the minimum data load or users, significant number of SBSs, though all under-utilized, are active. The saving potential of $85-90\%$ was calculated with implicit assumption of $0\%$ load \cite{Godor2012}.
\item The comparison between legacy and separation architecture is not fair in some studies. The separation architecture appears energy efficient due to other assumptions in the study rather than the separation itself. For example, a heterogeneous separated CRAN is energy efficient than a conventional HetNet mainly due to centralization of BBU (Baseband Unit) pool \cite{Liu2016}.
\item Similarly, the smaller consumption of a SBS compared to a macro BS is the main reason for separation architecture with small data cells to appear more energy efficient than legacy systems with only macro cells \cite{Wang2014}. 
\end{enumerate} 

We strongly emphasize that it is really important to reduce the carbon footprint of the communication industry which is expected to consume 20\% of all the world's electricity by 2025 (www.climatechangenews.com). However, it is important to do it correctly. The concept of providing coverage with minimal consumption of energy is important but some equipment should also be active to serve the little amount of data load always present in the system. The separation should not be between control signaling and data, it should be between control signaling plus data during low utilization phase and the rest of the data. Where exactly the split should take place, or how much data capacity should the low-power signaling infrastructure have, depends on the specific scenario, traffic, and power consumption of relevant equipment. 

The primary purpose of our analysis is to compare the relative merit of the separation architecture over legacy architecture in terms of energy saving potential. It is only one of the many performance measures for a mobile system. The merit of a system can only be judged against another if we consider the quality of service or quality of experience the system can provide along with the related costs, i.e., how efficient the system is in utilizing the resources to provide a certain grade of service \cite{Ge2015,Xiang2013}. The focus of our study remains, however, on the relative energy savings as it is the key reason to introduce the control-data separation architecture and the margins of savings should provide a motivation to investigate the other performance measures.

Generic assumptions about coverage, data traffic, energy management schemes, user demand, and power consumption of system elements are used to compute the total power consumption of different scenarios. Any complex handling of wireless channel effects, implementation of SBS {\it on/off} strategies, etc., are outside the scope of the present paper. Our assumptions, provided in sections \ref{scences} and \ref{the-model}, do not favor one scenario over another.  As salient contributions, this paper:
\begin{itemize}
    \item develops closed form expressions for activation probability of on-demand SBSs under different cellular system architectures and validates them using simulations;
    \item calculates energy consumption of a reference scenario across separation and non-separation architectures and studies the effects of parameter choices over relative performance;
    \item shows that the saving margins of separation architecture are not significant with in-depth analysis of alternative architectures.
\end{itemize}

The rest of the paper is organized as follows. We first outline the mobile architectures compared in this study in Section \ref{scences}, then introduce our analytical framework with a simple example in Section \ref{example-sec}. Section \ref{the-model} describes the system model, basic assumptions, and relevant measures from stochastic geometry. Section \ref{section_p_sat} calculates important probabilistic measures which are used in Section \ref{scenarios} to derive the activation probability of SBSs for each architectural alternative. Section \ref{results} discusses our numerical results in detail. Section \ref{rel-work} discusses related work and finally Section \ref{conc} summarizes the conclusions. 

\section{Mobile System Architectures}\label{scences}

\begin{figure}
\begin{center}
\includegraphics[scale = 0.43, trim=50 100 50 200, clip]{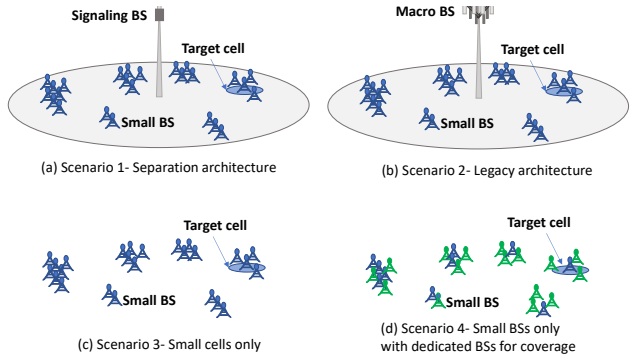}
\caption{Deployment scenarios in the study. Analysis is focused on {\it target cell}.}
\label{Scenarios_all}
\vspace{-5mm}
\end{center}
\end{figure}

Figure \ref{Scenarios_all} shows the four architectural alternatives, for mobile communication systems, compared in this study. Scenario 2, Fig.\ \ref{Scenarios_all}-(b), shows the legacy architecture, consisting of a typical macro cell along with clusters of homogeneous SBSs, deployed at the hot-spots to carry most of the data transmissions. We assume that small cells can be turned {\it off}, or put in {\it sleep} mode, if there is no data to send, although, the macro BS will remain {\it on} to cater for the control signaling along with data. Scenario 1, Fig.\ \ref{Scenarios_all}-(a), is similar to scenario 2, except that it has a signaling only macro BS \cite{Zaidi2014}, instead of a standard macro BS, to provide only always-{\it on} control signaling in a separation architecture. Signaling macro BS consumes less amount of energy as it does not require large antenna arrays and high data rate processing units and is the main reason behind separation architecture's energy efficiency. On the other hand, the BSs responsible for data transmission in a separation architecture are the same as the standard SBSs \cite{Ishii2012} and consume similar amount of energy.   
Scenarios 3 and 4, Fig.\ \ref{Scenarios_all}-(c),(d), are hypothetical scenarios with SBSs only to provide coverage and data services to all users. SBSs in scenario 3 are required to be {\it on} if there is data to send or area to cover. Scenario 4 assumes that some of the small cells are specially positioned for providing coverage and are always {\it on}, shown by green SBSs in Fig.\ \ref{Scenarios_all}-(d). All small cells in every scenario are assumed to be similar and except scenario 4's always-{\it on} cells, they can be turned {\it off} or put to {\it sleep} if not needed. 

A SBS in all of the scenarios can be a small eNodeB, or small gNB, or a RRH (Radio Remote Head) with small range in a CRAN setting. Similarly, a macro BS can be an eNodeB, or gNB, or a macro cell RRH. We select one small cell, called the {\it target cell}, in a hot-spot of each of the scenarios as shown in Fig.\ \ref{Scenarios_all} for our comparative study. The target cell is assumed to be overlapped by $N$ small cells in all scenarios. The measure to compare in our study is the energy consumed in a typical day by these $N+1$ small cells along with the relative consumption of the signaling or standard macro BSs in scenarios 1 and 2 respectively when serving similar load. 

\section{Case Study: An Example Scenario} \label{example-sec}

\begin{figure}
\begin{center}
\includegraphics[scale = 0.5, trim=75 120 250 230, clip]{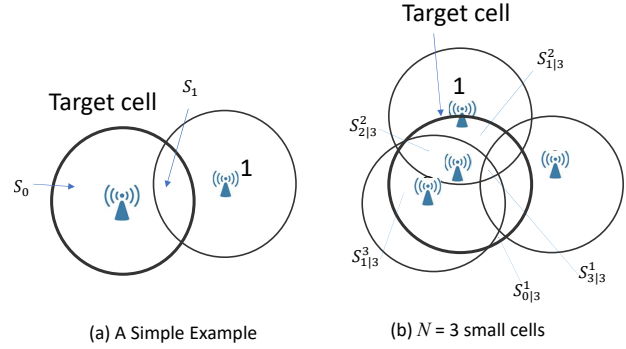}
\caption{Target cell with thicker boundary, and overlapping small cells, (a) $N= 1$, (b) $N=3$.}
\label{Target_cell}
\vspace{-5mm}
\end{center}
\end{figure}

Before formally describing the system model and developing the generic expressions for our study, we explain our approach with a simple example. In this example, we will compare two scenarios: (i) separation architecture (scenario 1), and (ii) small cells only (scenario 3). 

Let us consider two randomly deployed homogeneous small cells, each has area $A$, with overlapping coverage. One of these cells is the {\it target cell} and we will focus on determining its activation probability. The other identical cell is called cell 1, as shown in Fig.\ \ref{Target_cell}-(a). Both SBSs can either be {\it on}, with probability $\beta$ or {\it off} with probability $1-\beta$. The area of the target cell overlapped by cell 1 is denoted by $S_1$ and the rest of its coverage area by $S_0$. 

In scenario 1 of separation architecture, the small cells are responsible for only data transmissions and can be {\it on} if there is a user to serve. If cell 1 is {\it off}, any user in the area $A$ of the target cell will trigger the activation of the target cell. If we assume PPP (Poisson Point Process) spatially distributed users with density $\lambda$, then the activation probability of the target cell is $p\textsubscript{\it off} = 1-e^{-\lambda A}$ when cell 1 is {\it off}\footnote{Here, the subscript indicates the cell 1 state.}.

If BS 1 is {\it on}, the target BS will be {\it on} if there is at least one user in the area $S_0$, or the users in cell $1$ exceeds the {\it saturation capacity} of BS 1. Assume that BS 1 can serve maximum $N_{\rm small}$ number of users. If the area spanned by $S_0$ is $F_0A$, i.e., $F_0$ is the fraction of coverage area $A$ served only by the target cell, the probability of at least one user in $S_0$ is $
p\textsubscript{{\it on},0} = 1-e^{-\lambda F_0A}$. Similarly, the probability of users exceeding the saturation capacity of cell 1 is \vspace{-1mm}
\[p\textsubscript{{\it on},1} = \sum_{j=N_{\rm small}+ 1}^\infty \frac{(\lambda A)^j}{j!}e^{-\lambda A}.\]
The activation probability of the target cell in the case when cell 1 is {\it on} is the union of two independent events leading to $p\textsubscript{{\it on},0}$ and $p\textsubscript{{\it on},1}$, which can be given as\vspace{-2mm}
\[\vspace{-1mm}
p\textsubscript{\it on}
= 1 - e^{-\lambda F_0A}\sum_{j=0}^{N_{\rm small}} \frac{(\lambda A)^j}{j!}e^{-\lambda A}.
\]
Summing the probabilities of both mutually exclusive events when cell 1 is {\it on} and {\it off}, i.e., , $p\textsubscript{\it on}\beta_1$ and $p\textsubscript{\it off}(1-\beta_1)$, we get the activation probability of the target cell $\beta_1$\vspace{-1mm}
\begin{eqnarray}\label{beta-s-example}
\beta_1 &=& \beta_1\Bigl(1 - e^{-\lambda F_0A}\sum_{j=0}^{N_{\rm small}} \frac{(\lambda A)^j}{j!}e^{-\lambda A}\Bigr)\nonumber\\
&&+(1-\beta_1)\Bigl(1-e^{-\lambda A}\Bigr).
\end{eqnarray}
$\beta_1$ represents the activation probability of all small cells for scenario 1.

In scenario 3, we assume that no separate signaling macro cell is available and we only have SBSs. In this case, the SBSs should also provide coverage. If covering $1-\epsilon$ fraction of the area would be sufficient, where $\epsilon$ is an arbitrary number close to $0$, then the target cell should be activated if the uncovered fractional area exceeds $\epsilon$. When BS 1 is {\it off}, the target BS should be {\it on} as $100\%$ of the cell area is uncovered. When BS 1 is {\it on}, the target BS will be activated if $F_0 > \epsilon$, or if users in cell $1$ exceed $N_{\rm small}$. We can rewrite \eqref{beta-s-example} for $\beta_3$, the activation probability of the target cell in scenario 3, as\vspace{-1mm} 
\begin{eqnarray}\label{beta-ns-example}
\beta_3 = \beta_3\Bigl(1 - \Pr(F_0 < \epsilon)\sum_{j=0}^{N_{\rm small}} \frac{(\lambda A)^j}{j!}e^{-\lambda A}\Bigr)
+(1-\beta_3).
\end{eqnarray}
We will use the results from stochastic geometry, described in the next section, to generalize the model for $N$ small cells, realistic traffic models, and also for the rest of two scenarios depicted in Fig.\ \ref{Scenarios_all}. $\lambda$ can be obtained from daily traffic profiles \cite{ITU2015} to calculate activation probabilities, which can then be used with the power consumption of associated equipment to determine the total daily consumption for each scenario.
\vspace{-1mm}
\section{Model Details and Key System Parameters}\label{the-model}

\subsection{Spatial Network Deployment and User Distribution}

We consider a cluster of $N$ homogeneous SBSs with overlapping coverage areas to a similar small cell called the {\it target cell} as shown in Fig.\ \ref{Scenarios_all} and illustrated in detail in Fig.\ \ref{Target_cell}. We will calculate the activation probability of the target cell considering all possible events associated with users' presence in the target cell and the active/inactive states of the overlapping $N$ small cells. A Poisson Cluster Process (PCP) \cite{Saha2017} is used for the spatial distribution of SBSs. In a PCP, a parent PPP determines the locations of $N_c$ hot-spots or clusters in a macro cell, and $N$ SBSs are uniformly distributed within each cluster. 

Hexagonal grid models are also commonly used both in industry and academia for system-level performance evaluation of wireless cellular networks \cite{3GPP06, Rappaport02, Viterbi91}. The previous papers such as \cite{Lee2013,Andrews2011} showed that PPP models are as good as (and even better in some deployment scenarios than) popular hexagonal grid models for modeling BS locations in terms of coverage probability estimation when compared with real-world BS deployments. In particular, the PPP based location model leads to a lower performance bound, whilst the grid model results in an upper performance bound with almost the same deviations from the experimental data \cite{Andrews2011}.  It was also observed that PPP models become even more suitable for networks with femto-cells which may take up unknown and unplanned positions \cite{Dhillon2012}. Besides its modeling accuracy, the PPP model in this work provides us with analytical tractability to derive network energy efficiency formulas for different BS deployment scenarios, which are otherwise prohibitively hard to derive, as in many other previous work in the field adopting PPPs to characterise BS locations \cite{Chat2018,Fereydooni2018,Yang2018,Sun2018,Ghatak2019,Bursa2019,Liu2019,Mahbas2019,Yang2019}. The other point processes providing a better fit to real-world BS deployment data than PPPs and grid models (but at the expense of losing analytical tractability) are hard-core processes (HCPs), Strauss process and Geyer saturation process \cite{Riihijarvi10, Guo2013, Taylor12}.\footnote{The goodness-of-fit can be determined by the selected pseudo-likelihood functions as well as point process statistics such as empty space function, area of Voronoi cells, J function and L function.} We also test the robustness of our results obtained for PPPs using simulation experiments with Mat\'ern I HCPs \cite{Elsawy2013} that impose minimum distance restriction between BSs. The energy saving predictions obtained by PPPs continue to hold the same with those obtained by using HCPs.

The users are also assumed to be spatially distributed according to a PPP with density $\lambda$ within the hot-spots. Their respective traffic demand follows {\it iid} self-similar (Pareto inter-arrivals) or Poisson distribution. It is assumed that users can select any reachable BS randomly with equal probability causing possible activation for inactive SBSs. SBSs can either be {\it on}, with probability $\beta$ or {\it off} with probability $1-\beta$. Section VI describes the methods to calculate these activation probabilities in detail, which in turn determine the steady state distribution for the SBS activation process under different architectural choices, and Section \ref{section_valid} verifies the methods by simulation experiments over range of model parameter values. 

Circular cell models are more suitable for SBSs where line-of-sight (LOS) communication is more likely and when the users are considered out of the range if the associate received power is lower than a threshold \cite{Ding2018}. Moreover, the transmission power almost remains the same for all users within a small cell.  We used circular small cells in Fig. \ref{Target_cell} and in our model validation experiments but it is not a requirement of the analysis model. It is required, however, that $N$ cells overlap the target cell, and all have the same area and perimeter. For example, in case of circular coverage area with radius $R$, any SBS situated within $2R$ distance from the target SBS will have overlapping coverage with the target cell.
\vspace{-2mm}
\subsection{$k$-Coverage Probability}

For $N$ homogeneous SBSs distributed according to the PPP with density $\lambda$, the areas indicated by $S_{k|N}^i$ in Fig,\ \ref{Target_cell}-(b), where $k = 0, 1, \cdots, N$ and $i = 1, \cdots, {N\choose k}$, are the spaces covered by exactly $k$ SBSs along with the target cell. The figure does not label all spaces for the sake of illustration clarity. The probability that a point $X$ belongs to one of the $S_{k|N}^i$ \cite{Lazos2006, Santalo2004} is 
\[ \Pr(X \in S_{k|N}^i) = \frac {(2 \pi A)^k (2 \pi A + P^2)^{N-k}}{(2 \pi (2A) + P^2)^N},\] 
where $A$ is the area of each of the small cells and $P$ is their perimeter. The probability that a point $X$ would be covered by exactly $k$ SBSs besides target cell, i.e., $X$ belongs to $S_{k|N}$, where $S_{k|N} = \cup_{i=1}^{N\choose k} S_{k|N}^i$, is \[\Pr(X \in S_{k|N}) = {N \choose k } \Pr(X \in S_{k|N}^i).\] Interestingly, $\Pr(X \in S_{k|N}^i)$ is also equal to the fraction of the area of target cell spanned by $S_{k|N}^i$ \cite{Lazos2006, Santalo2004}, i.e., if we denote $f_{k|N}^i$ as the fractional area of $S_{k|N}^i$ then
\begin{eqnarray}
f_{k|N}^i = \Pr(X \in S_ {k|N}^i) 
= \frac{(2 \pi A)^k (2 \pi A + P^2)^{N-k}}{(2 \pi (2A) + P^2)^N}. \label{f-k-i}
\end{eqnarray}     
Similarly,
\begin{eqnarray}
f_{k|N}\hspace{-1mm} =\hspace{-1mm}\ \Pr(X \in S_{k|N})\hspace{-1mm} =\hspace{-1mm} {N \choose k } \frac{(2 \pi A)^k (2 \pi A + P^2)^{N-k}}{(2 \pi (2A) + P^2)^N},\label{f-k}
\end{eqnarray}
where $f_{k|N}$ is the fraction of the area of the target cell occupied by $S_{k|N}$. $f_{k|N}^i$ and $f_{k|N}$ can take values from $0$ to $1$. Detailed relevant proofs can be accessed from \cite{Lazos2006} and \cite{Santalo2004}.
\vspace{-2mm}
\subsection{Traffic Model}\label{section_traffic}

We assume that $Y_i \in \{0, 1, 2, \cdots\}$ represents the traffic demand of $i$th user in terms of number of Resource Blocks (RBs) for time interval $(0, \tau]$. Here, $\tau$ can be the length of one radio frame in 5G, whose typical value  is $10$ms. Each radio frame can have maximum number of $140-2240$ symbols or RBs depending on the traffic configuration of 5G\footnote{https://www.sharetechnote.com/html/5G/5G\_FrameStructure.html}. 
Our traffic model generates the demand for each user in the area of interest in each time interval $\tau$ using the model parameters.

Specifically, $Y_i$'s are assumed to be {\it iid} random variables. For Poisson distributed traffic with arrival rate $\mu$, $P_{Y_i}(n)$, i.e., the probability that user $i$ needs to send or receive $n$ RBs in $(0, \tau]$, is given as\vspace{-1mm}
\begin{eqnarray}
P_{Y_i}(n) = \frac{(\mu \tau)^n}{n!}e^{-\mu \tau}.
\end{eqnarray}

We also consider self-similar traffic with heavy-tailed distribution as it provides more realistic representation of modern data traffic \cite{Xiang2013}. A popular example of self-similar traffic is Pareto distributed inter-arrival times, $T$, i.e.,$f_T(t) = \frac{a b^a}{t^{a+1}}$, where $b, 0<b\leq t,$ is the minimum possible value for $t$, and $a, a \in (1, 2],$ reflects the heaviness of the distribution tail \cite{Xiang2013}. In this case, $P_{Y_i}(n)$, i.e., the probability that user $i$ needs to send or receive $n$ RBs in $(0, \tau]$, is \cite{Singhai07}\vspace{-1mm}
\begin{eqnarray}
P_{Y_i}(n) = \frac{(a \log \tau/b)^n}{(\tau/b)^a n!},~\forall i.
\end{eqnarray}

More technically, the collection of points $\left(X_i, Y_i\right)_{i = 1}^\infty$, with $X_i \in \mathbb{R}^2$ representing the location of the $i$th user, form a {\it Marked Poisson Process}  in $\mathbb{R}^3$ \cite{Baccelli10}. Here, $Y_i$ is the mark associated with user $i$'s location. The probability that $j$ users generate a combined traffic demand of $n$ RBs in cell area $A$ can be computed using convolution of independent random variables.
\begin{eqnarray}
\hspace{-1mm}\Pr\left(\sum_{i=0}^j Y_i = n\right)\hspace{-1mm}
=\hspace{-1mm}\left\{ \hspace{-1mm}\begin{array}{ll}
\hspace{-1mm}\frac {(\lambda A)^j}{j!}e^{-\lambda A} \frac{(j\mu \tau)^{n}}{n!}e^{-j\mu \tau} \hspace{-2mm}&\hspace{-2mm} \mbox{Poisson}\\
\hspace{-1mm}\frac {(\lambda A)^j}{j!}e^{-\lambda A} \frac{(ja\log{\tau/b})^{n}}{(\tau/b)^{ja}n!} \hspace{-2mm}&\hspace{-2mm} \mbox{Self-similar}
\end{array}\right.\hspace{-2mm},\hspace{-2mm}
\label{prob_traffic}
\end{eqnarray}
where $\lambda$ is the user density per unit area. The mean traffic demand from an area $A$ in $(0, \tau]$ is $\lambda A\mu \tau$ in case of Poisson traffic and $\lambda A a\log(\tau/b)$ in case of self-similar traffic.
\vspace{-3mm}
\subsection{Saturation Capacity of a BS}\label{Section_saturation}

In this part, we will provide two important definitions that will be instrumental in determining BS activation probabilities in Section \ref{scenarios} by identifying their measure of communication capacity.
\vspace{-1mm}
\begin{definition}
Saturation capacity $N_{\rm small}$ of a SBS is defined as the maximum number of RBs it can allocate to users within its coverage area. Similarly, $N_{\rm macro}$ is the maximum number of RBs that a macro BS can allocate to the users within its coverage area.
\end{definition}
\vspace{-2mm}

We also define another term called {\it relative saturation capacity}. This is a useful metric that will be used to characterize the relative fraction of RBs allocated to a set of users from a particular part of the cell and contributing towards the saturation of the respective BS.
\vspace{-2mm}
\begin{definition}
The {\it relative saturation capacity} $N_{f}$ is the maximum number of RBs that can be served by a cell in a fractional area $f$. Using the symbol $\lfloor x \rceil$ for rounding $x$ to the nearest integer, $N_f$ is set to $N_{f} =\lfloor fN_{\rm cell} \rceil$, where $N_{\rm cell}$ is either $N_{\rm macro}$ or to $N_{\rm small}$ depending on the BS type.
\end{definition}

As described above, $Y_i$ is the traffic demand of each user $i$ with position $X_i$. When a BS is in saturation, we have $\sum_{j: X_j \in S} Y_j = N_{\rm cell}$, where $S$ represents the associated cell with saturation capacity of $N_{\rm cell}$. The relative saturation capacity measure defined above is indeed the conditional expected number of RBs (up to a rounding error) served by a BS in a particular area of interest $S^* \subseteq S$, when the BS is in saturation, i.e.,
\begin{eqnarray*}
N_f = E\left[\sum_{i: X_i \in S^*} Y_i\right| \left. \sum_{j: X_j \in S} Y_j = N_{\rm cell}\right]. 
\end{eqnarray*}

In a PPP, users are uniformly distributed over a region given they are known to lie in this region. Thus, the fractional area associated with $S^*$ is also equal to the probability of finding $X_i$ in $S^*$ once it is known to be in $S$. Using this observation, $N_f$ is equal to $N_{\rm cell} f$ since this identity holds for any number of users in $S$ greater than zero.

The major issue resolved by the relative saturation capacity metric is to specify the number of users that can be allowed in the region associated with $f_{0|N}$ without activating the target SBS. For example, in scenario 1, the target SBS will be activated when there is any user in $f_{0|N}$ and in scenario 3, coverage to this area is a more dominant issue, as shown in Section \ref{example-sec}. On the other hand, in scenarios 2 and 4, we have a macro BS and always-{\it on} SBSs covering this area respectively and some traffic can be allowed in this region without activating the target cell. The important question here is how much traffic from this region can be accommodated without triggering activation of the target cell. The concept of relative saturation capacity allows us to quantify the traffic demand associated with $f_{0|N}$ without activating the target cell as $\lfloor \frac{N_{\rm macro}}{N_cA_c}f_{0|N}A \rceil$ in scenario 2 and $\lfloor\eta N_{\rm small}f_{0|N} \rceil$ in scenario 4. Here, $N_c$ is the number of clusters or hot-spots of area $A_c$ in a macro cell and $\eta$ compensates for possible overlapping of always-{\it on} cells. Details are given in Theorems \ref{beta-nsm-th} and \ref{beta-ns3-th}.
\vspace{-3mm}
\subsection{Power Consumption Model}\label{PC-levels}

Our power consumption model uses the number of active BSs, their types (i.e., SBSs and macro cell BSs), and realistic power consumption values for each BS type as high-level system parameters to compute power consumption of four different network architectures. This approach, albeit simplicity of our power consumption model, will lead to insightful analytical expressions for energy consumption comparison.

\subsubsection{Macro BS Power Consumption}

We assume that a macro BS consumes maximum power $P_{\rm macro}$ throughout the day. Out of the total power consumption of a macro BS, $55\% - 60\%$ is consumed in the power amplifier and is load dependent \cite{Auer2011}. Hence, a more realistic power consumption model for macro BSs can take network load, transmission rates, channel state, and multiple-access interference into account to optimize the power consumption. However, 
in a load-based model, macro BS power consumption will vary depending on the network load and could be less than the maximum consumption, whereas the energy consumption of SBSs will mostly remain independent of the load as shown in \cite{Auer2011}. This will further decrease the gap between power consumption of the legacy architecture and that of separation architecture.

\subsubsection{Signaling BS Power Consumption}

The power consumption of signaling only macro BS will be taken as $P_{\rm sig-macro} = P_{\rm macro}/\rho_1$ for $\rho_1 \geq 1$. In practice, $\rho_1$ can be taken as $6$ since the signaling BS requires single antenna reducing the consumption by $1/3$, and low rate data processing, reducing the consumption by $1/2$ by minimizing the load-dependent fraction \cite{Zaidi2015}. In a CRAN setting, the power consumption of a signaling only RRH with relevant BBU can be taken as $P_{\rm macro}/\rho_2$ for $\rho_2 \approx 18$. On top of single antenna use and low load, a RRH does not need any cooling component either, reducing a further $1/3^{\text{rd}}$ of the power. We vary these ratios in Section \ref{results} to ascertain that our conclusions are not sensitive to the parameters' values. 

\subsubsection{SBS Power Consumption}

The power consumption of a SBS is almost independent of load \cite{Auer2011}. Moreover, since the communication is over short distances and mostly LOS \cite{Ding2018}, the transmission power is not significantly different from one user to another within the cell. We take the power consumption of a SBS, as $P_{\rm small} = P_{\rm macro}/\rho_3$. $\rho_3$ can  be taken as $100$ \cite{Auer2011}. In a CRAN, small RRHs' consumption is $P_{\rm small}/\rho_4$. A realistic value for $\rho_4$ could be $1.15$ due to half (power amplifier) consumption \cite{Alhumaima2017}. The power consumption in {\it sleep} mode is selected as $P_{\rm sleep} = 0.1$W in our simulation study \cite{Filippini2017}. The implementation of {\it on/off/sleep} cycles has many issues \cite{Feng2017} but it is outside the scope of the present paper. However, we assumed $P\textsubscript{\it off} = P_{\rm sleep}$ in order to avoid over-estimation of energy savings\footnote{Over-estimation of energy savings is equivalent to under-estimation of energy consumption.}. 
\vspace{-3mm}
\section{Saturation Probabilities}\label{section_p_sat}

As discussed in Section \ref{example-sec}, we need the saturation probability of the overlapping BSs, i.e., the probability that they are within their saturation capacity, in order to calculate the activation probability of target BS. A user can be served by either of the reachable BS and the thinning theorem can be used to calculate the probability of users assigned to each of the cells.

\begin{theorem}\label{P-sat}
The probability for each active SBS among $l$, $0 \leq l \leq N$, to remain within the saturation limit $N_{\rm small}$, is given by \vspace{-2mm}
\begin{eqnarray}
P_{ss}\hspace{-1mm} =\hspace{-1mm} \sum_{j=0}^{\infty} \frac{ (\lambda f(l\hspace{-1mm}-\hspace{-1mm}1,0)A)^j} {j!} e^{-\lambda f(l-1,0)A}\hspace{-1mm}\sum_{n=0}^{N_{\rm small}}\hspace{-1mm} \hspace{-0.8mm}\Pr\hspace{-1mm}\left(\hspace{-0.3mm}\sum_{i=0}^jY_i\hspace{-1mm} = \hspace{-1mm}n\hspace{-1mm}\right),\hspace{-2mm}
\label{Prob-sat}
\end{eqnarray}
where $\Pr(\sum_{i=0}^jY_i = n)$ is given in \eqref{prob_traffic} and \[f(l-1,0) = \sum_{k = 0}^{l-1} {{l - 1} \choose {k}} \frac{f_{k|l-1}^i}{k+1}.\]
\end{theorem}
\begin{proof}
Each active small cell will have $l-1$ overlapping active cells. In $f_{k|l-1}^i$, any BS from $k$ and the active cell can be selected with equal probability to serve a user. As a consequence, the average number of users served by each BS in $f_{k|l-1}^i$ is $\lambda/(k+1)$ by using the thinning theorem for Poisson processes. Hence, average number of users served by each BS is given by the Poisson random variable with mean 
\begin{eqnarray*}
\sum_{k=0}^{l-1} \frac{\lambda}{k+1}{{l-1} \choose {k}} f_{k|l-1}^i A &=& \lambda A  \sum_{k = 0}^{l-1} {{l - 1} \choose {k}} \frac{f_{k|l-1}^i}{k+1}\\
&=& \lambda A f(l-1,0).
\end{eqnarray*} \vspace{-3mm}
\end{proof}

One important remark about \eqref{Prob-sat} is that $\sum_{n=0}^{N_{\rm small}} \Pr\left(\sum_{i=0}^jY_i = n\right) = \Pr\left( \sum_{i=0}^jY_i \leq N_{\rm small} \right)$. We prefer to provide an expression for $P_{ss}$ as given in \eqref{Prob-sat} because we can analytically express the probability $\Pr\left(\sum_{i=0}^jY_i \leq N_{\rm small} \right)$ in closed form  for all $n \geq 0$.   

\begin{theorem} \label{P-sat2}
Consider the scenario in which there are $M$ always-on small or macro cells and $l$ active SBSs. Assume $M$ always-on cells collectively cover the area of these $l$ active small cells. Let $A_m \in [0, 1]$, for $m=1, \ldots, M$, be the fraction of area of any of these $l$ active small cells simultaneously covered by $m$ always-on cells (i.e., any point in such a region also belongs to $m$ always-on cells). If users are assigned to any active SBS in range with equal probability, then the probability that each of the $l$ SBSs to remain within the saturation limit $N_{\rm small}$ can be expressed by  
\begin{eqnarray}
P_{ss_2}(M) = \sum_{j=0}^{\infty}\ddfrac{ (\lambda A \sum_{m=1}^M A_m f(l-1,m))^j} {j!} e^{-\lambda A \sum_{m=1}^M A_mf(l-1,m)}\Pr\left( \sum_{i=0}^jY_i \leq N_{\rm small} \right),
\label{Prob-sat2}
\end{eqnarray}
where $\Pr\left(\sum_{i=0}^jY_i = n\right)$ is given in \eqref{prob_traffic}, $\sum_{m=1}^M A_m = 1$ and \[
f(l-1,m) = \sum_{k = 0}^{l-1} {{l - 1} \choose {k}} \frac{f_{k|l-1}^i}{k+1+m}.\] 
\end{theorem}
\begin{proof}
The area of a given small cell $A$ can be expressed as $A = \sum_{m=1}^M A_mA$, where $A_m \in [0, 1]$ and $\sum_{m=1}^M A_m = 1$. Each $A_m$ is the fractional area where $m$ always-{\it on} cells overlap, and users can be assigned to any reachable BS with equal probability. Using the thinning theorem, the Poisson coefficient seen by each BS among $l$ active SBSs is
\begin{eqnarray*}
\sum_{k=0}^{l-1} {{l-1} \choose {k}} f_{k|l-1}^i \sum_{m=1}^M\frac{\lambda}{k+1+m}A_mA 
&=& \lambda A \sum_{m=1}^M A_m\sum_{k = 0}^{l-1} {{l - 1} \choose {k}} \frac{f_{k|l-1}^i}{k+1+m}
\\&=& \lambda A \sum_{m=1}^M A_m f(l-1,m).
\end{eqnarray*}
\end{proof}

\begin{corollary}\label{P-sat-on}
Consider the scenario where $M$ always-{\it on} small cells collectively cover the area of $l$ active small cells as described in Theorem \ref{P-sat2}. For each of the $M$ special always-{\it on} SBSs, the probability to remain within the saturation limit $N_{\rm small}$ is given as 
\begin{eqnarray}
P_{ss-on}(M) = \sum_{j=0}^{\infty}\frac{ (\lambda A \sum_{m = 0}^{M-1} A_m f(l,m))^j} {j!} e^{-\lambda A \sum_{m=0}^{M-1} A_mf(l,m)}\Pr\left( \sum_{i=0}^jY_i \leq N_{\rm small} \right),
\label{Prob-sat-on}
\end{eqnarray}
where $A_m$ is the fractional area where $m$ always-{\it on} cells overlap, $A_m \in [0,1]$ and $\sum_{m=0}^{M-1} A_m = 1$, and $\Pr(\sum_{i=0}^jY_i = n)$ is given in \eqref{prob_traffic} and \vspace{-2mm} \[
f(l,m) = \sum_{k = 0}^{l} {{l} \choose {k}} \frac{f_{k|l}^i}{k+1+m}, ~ m = 0, 1, \cdots, M-1.
\]
\end{corollary}
\begin{proof}
The proof is similar to the one given for Theorem 2. It is omitted to avoid repetitions.  
\end{proof}

\begin{corollary}\label{P-satm}
The probability, $P_{sm}$, for macro cell to remain within its saturation limit $N_{\rm macro}$ can be given according to
\begin{eqnarray}
P_{sm} =
\sum_{N_c=1}^{\infty}\frac{ (\bar{N_c})^{N_c}e^{-\bar{N_c}}} {N_c!(1-e^{-\bar{N_c}})} 
\sum_{j=0}^{\infty} \frac{ (\lambda f(l,0)N_cA_c)^j} {j!} e^{-\lambda f(l,0)N_cA_c}
\Pr\left( \sum_{i=0}^jY_i \leq N_{\rm macro} \right)\label{Prob-satm}
\end{eqnarray}
in the presence of $l$ small active cells in each hot-spot of area $A_c$ and perimeter $P_c$, where $\bar{N_c}$ is the mean number of clusters in a macro cell and $\Pr(\sum_{i=0}^jY_i = n)$ is given in \eqref{prob_traffic}. Also, 
\begin{eqnarray*}
f(l,0) = \sum_{k = 0}^{l} {{l} \choose {k}} \frac{f_{k|l}^i}{k+1}, ~\mbox{where}~
f_{k|l}^i = \frac{(2 \pi A)^k (2 \pi A_c + PP_c)^{l-k}}{(2 \pi (A + A_c) + PP_c)^l}.
\end{eqnarray*}
\end{corollary}

\begin{proof}
When a macro cell is overlapped by $l$ SBSs in each hot-spot, the number of users served by the macro BS per hot-spot is given by a Poisson random variable with mean $\lambda f(l,0)A_c$, similar to Theorem \ref{P-sat}. Here, $f_{k|l}^i$ reflects the fractional area of exact $k$-overlap when $l$ small cells of area $A$ and perimeter $P$ are randomly deployed over a hot-spot of area $A_c$ and perimeter $P_c$ \cite{Lazos2006,Santalo2004}. This expression for $f_{k|l}^i$ is used only for calculation of $P_{sm}$. For all other probabilities, $f_{k|l}^i$ is calculated by \eqref{f-k-i}. Summing over $N_c$ identical and independent clusters, we obtain a Poisson random variable with mean $\lambda f(l,0)N_cA_c$, conditioned on $N_c$, for the total number of users covered by the macro cell. $N_c = 1, 2, \cdots$ is also a Poisson random variable itself in the PCP model \cite{Saha2017} and \eqref{Prob-satm} is achieved when the probability of attaining $N_{\rm macro}$ RBs in a macro cell for specific $N_c$ is averaged over the distribution of $N_c$.
\end{proof}

\section{Activation Probabilities}\label{scenarios}

In this section, we develop the activation probabilities for target cell in each of the architectural alternatives discussed in Section \ref{scences}.
\vspace{-1mm}
\subsection{Scenario 1: Separation Architecture}\label{wSig}

In this scenario, we assume that a signaling macro cell is overlaying the target cell and $N$ small cells. A signaling macro cell \cite{Zaidi2014} is a macro cell which is responsible to provide low-rate control signaling only. In this case, $N$ SBSs and the target SBS will be {\it on} only if there is a user to serve. We define $\beta_1$ as the activation probability of target and $N$ SBSs in scenario 1. 

\begin{theorem}\label{beta-s-th}
The activation probability $\beta_1$ of the target cell, also for each of the $N$ randomly deployed cells, for scenario 1 is given as\vspace{-1mm}
\begin{eqnarray}
\beta_1 = \sum_{l = 0}^N \beta_1^l {(1-\beta_1)}^{(N-l)} \left (1- p_0(P_{ss})^l \right ),\label{beta-s}
\end{eqnarray} 
where $P_{ss}$ is given in Theorem \ref{P-sat} and $p_0$ is given according to 
\begin{eqnarray*}
p_0 = \Pr(\text{no traffic in}~f_{0|l}A) =
\left\{ \begin{array}{ll}
 e^{-\lambda f_{0|l}A(1-e^{-\mu\tau})} & \mbox{for Poisson traffic}\\
 e^{-\lambda f_{0|l}A(1-(b/\tau)^a)}& \mbox{for self-similar traffic}
\end{array}\right..
\end{eqnarray*}
\end{theorem}\vspace{-1mm}
\begin{proof}
Consider the case in which there are $l$ active cells out of $N$ SBSs. In this case, the target cell will not be turned {\it on} if and only if there is no traffic in the fractional area $f_{0|l}$ covered only by the target cell and all $l$ active cells remain unsaturated. The probability for the former event is equal to $p_0$, which is given by (cf.\ \eqref{prob_traffic})
\begin{eqnarray*}
p_0 = \Pr\left(\text{no traffic in}~f_{0|l}A \right) = \sum_{j=0}^{\infty} \frac{ (\lambda f_{0|l}A)^j} {j!} e^{-\lambda f_{0|l}A} \Pr\left(\sum_{i=0}^jY_i = 0\right). 
\end{eqnarray*}

On the other hand, the probability for the latter event is $(P_{ss})^l$. Using symmetry in the problem and summing up all possible cases, we arrive at \eqref{beta-s}, which concludes the proof. 
\end{proof}

We note that \eqref{beta-s} in Theorem \ref{beta-s-th} needs to be solved numerically to obtain the activation probability $\beta_1$ in this network deployment scenario. It is shown in validation experiments in Section \ref{section_valid} that this numerical calculation can be carried out efficiently and $\beta_1$ matches the simulated activation values accurately. 


\subsection{Scenario 2: Legacy Architecture}

In this scenario, a standard macro cell overlay the target cell and $N$ overlapping small cells catering for coverage as well as for some of the data transmissions. 

\begin{theorem} \label{beta-nsm-th}
Assume that users can be assigned randomly to an overlaying macro cell and any $l$ active cells. Then, the probability of activation of target SBS, $\beta_2$, is given by
\begin{eqnarray}
\beta_2= \sum_{l = 0}^N \beta_2^l {(1-\beta_2)}^{(N-l)} 
\hspace{-1mm}\left (1- p_{0_2} (P_{sm}(P_{ss_2}(1))^l + (1-P_{sm})(P_{ss})^l) \right),\hspace{-3mm}\label{beta-nsm}
\end{eqnarray} 
where $p_{0_2} = \Pr(\text{no traffic than that served by macro BS in }f_{0|l}A)$, $P_{ss}$, $P_{ss_2}(1)$, and $P_{sm}$ are given in \eqref{Prob-sat}, \eqref{Prob-sat2} and \eqref{Prob-satm}, respectively.  
\end{theorem}
\begin{proof}
Similar to the proof of Theorem \ref{beta-s-th}, the target SBS will not be turned {\it on} in this scenario of $l$ active overlapping SBSs and a macro BS if and only if the traffic in $f_{0|l}A$ does not exceed the capacity that can be assigned to macro BS, and all macro BS and $l$ active SBSs remain unsaturated. The probability of the former event is
\[p_{0_2} = \sum_{j=0}^\infty\frac {(\lambda f_{0|l}A)^j} {j!} e^{-\lambda f_{0|l}A}\Pr\left( \sum_{i=0}^jY_i \leq N_{0_2} \right),\] 
where $N_{0_2}$ represents the maximum number of RBs that can be allocated to the macro BS and associated with the users in $f_{0|l}A$. 
The probability of the latter event is the union of two mutually exclusive events. When the macro BS is within its saturation capacity, the users can be randomly assigned to the macro BS and $l$ SBSs. In this case, the probability that all active BSs remain unsaturated is $P_{sm}(P_{ss_2}(1))^l$. On the other hand, when the macro cell is saturated, the users can only be assigned to $l$ active cells and the probability that they remain unsaturated while the macro BS is saturated is $(1-P_{sm})(P_{ss})^l$. Using symmetry in the problem and summing up over all possible cases, we arrive at \eqref{beta-nsm}, which concludes the proof. 
\end{proof}\vspace{-2mm}

We note that we cannot provide a closed form analytical expression for $p_{0_2}$ in Theorem \ref{beta-nsm-th}, mainly due to unavailability of an accurate characterization of $N_{0_2}$. As one way of approximating $p_{0_2}$, we propose to use the approximation $N_{0_2} \approx {\lfloor \frac{N_{\rm macro}}{N_c A_c} f_{0|l}A\rceil}$, where $N_{\rm macro}$ is the saturation capacity of a macro BS and $N_c$ is the Poisson random variable representing number of hot-spots, of area $A_c$, in the macro cell. This approximation is also discussed in Section \ref{the-model}. By using this approximation for $N_{0_2}$, the probability $p_{0_2}$ can be approximated as follows
\[p_{0_2} \approx \sum_{j=0}^\infty\frac {(\lambda f_{0|l}A)^j} {j!} e^{-\lambda f_{0|l}A}\sum_{N_c =1}^\infty \frac{ (\bar{N_c})^{N_c}e^{-\bar{N_c}}} {N_c!(1-e^{-\bar{N_c}})} \Pr\left( \sum_{i=0}^jY_i \leq {\left\lfloor \frac{N_{\rm macro}} {N_cA_c}f_{0|l}A\right\rceil} \right).\] 

The derivation is similar to the one given for Corollary \ref{P-satm}, and hence omitted to avoid repetitions.  

\subsection{Scenario 3: Small Cells Only}

In this scenario, the SBSs provide control signaling as well as data transmissions. In order to provide coverage over $(1-\epsilon)$ of the total area, one or multiple SBSs should be {\it on} all the time. Also, as in scenario 1, an inactive SBS is turned {\it on} to serve users in case an active SBS is saturated, though, it can still provide control signaling. 

More specifically, for $l$, $0 \leq l \leq N$, active SBSs, the target cell will be turned {\it on} if the fractional area associated with $S_{0|l}$ is greater than $\epsilon$ in scenario 3. The stochastic geometry expressions given in \eqref{f-k-i} and \eqref{f-k} were utilized in our analysis above to derive the activation probabilities in scenarios 1 and 2 by calculating the mean values of fractional areas \cite{Santalo2004}. However, in this case, we also need to know the distribution of the random fractional area spanned by $S_{0 |l}$ to derive activation probabilities. To this end, we define a random variable $F_{0 |l} \in [0, 1]$, having mean $f_{0|l}$, which represents the fractional area spanned by $S_{0 |l}$ in an instantaneous deployment. We will use the probability $\Pr\left(F_{0 |l} < \epsilon \right)$ in the statement of Theorem \ref{beta-ns-th} to characterize the activation probability $\beta_3$ for the target cell in scenario 3. After the proof of this theorem, we will provide an efficient approach to approximate this probability.   

\begin{theorem}\label{beta-ns-th}
The activation probability $\beta_3$ of the target cell, also for each of the $N$ randomly deployed cells, for scenario 3 is given as\vspace{-2mm}
\begin{eqnarray}\vspace{-2mm}
\beta_3 = \sum\limits_{l = 0}^N \beta_3^l {(1-\beta_3)}^{(N-l)} \left(1-\Pr(F_{0|l} < \epsilon)(P_{ss})^l\right),\label{beta-ns}
\end{eqnarray} 
where $P_{ss}$ is given in Theorem \ref{P-sat} and $F_{0|l}$ is the random fractional area spanned by $S_{0|l}$.
\end{theorem}
\begin{proof}
For the case where there are $l$, $0 \leq l \leq N$, active cells, the target cell will not be turned {\it on} if $F_{0|l} < \epsilon$ and all $l$ SBSs remain within their saturation limit as in Theorem \ref{beta-s-th}. Using symmetry in the problem and summing up all possible cases, we arrive at \eqref{beta-ns}, which concludes the proof. \end{proof}


Similar to Theorem \ref{beta-nsm-th}, we cannot accurately characterize the probability expression $\Pr(F_{0|l} < \epsilon)$ in Theorem \ref{beta-ns-th}. There are some important results in the stochastic geometry literature, such as \cite{Moller2004}, which show that generalized geometric contents of many closed sets constructed over PPPs have Gamma type distributions. Poisson-Voronoi tessellation is an important example of such content. According to our search, however, there is no work relevant to the distribution of $F_{0|l}$. We still find that the assumption of $F_{0|l}$ having a Gamma distribution is a workable choice. This assumption lets us calculate the coverage probabilities with reasonable accuracy, as shown by the model validation experiments in Section \ref{section_valid}. We also simulated scenarios with random deployments of SBSs, calculated empirical distributions of $F_{0|l}$, and calculated $p$-values for distribution fit for relevant Gamma distributions. The calculated $p$-values provide strong statistical significance for most of the cases. 

In order to provide an approximation for $\Pr(F_{0|l} < \epsilon)$, let us define $g(x)$ as the probability density function (pdf) of $F_{0|l}$. We set $g(x) = \Gamma(x; \alpha,\theta)/C$, where $x \in [0, 1]$ and $C$ is the normalizing constant, $\alpha$ is the shape parameter and $\theta$ is the scale parameter of Gamma distribution. $g(x)$ is then given as\vspace{-0mm}
\begin{eqnarray*}
g(x) =  \frac{{x}^{\alpha-1}e^{-x/\theta}}{C{\theta}^{\alpha} \Gamma(\alpha)},
\mbox{where}~\Gamma(\alpha) = \int_0^\infty x^{\alpha-1}e^{-x}dx  
\mbox{ and }
C = \int\limits_0^1 \frac{{x}^{\alpha-1}e^{-x/\theta}}{{\theta}^{\alpha} \Gamma(\alpha)} dx.
\end{eqnarray*}
Using the properties of Gamma pdfs, the parameters $\alpha$ and $\theta$ can be calculated through mean and variance. In particular, the mean of $F_{0|l}$ is equal to $f_{0|l}$, which can be calculated by using \eqref{f-k}. Let $\sigma_{0|l}$ be the standard deviation of $F_{0|l}$. Then $\theta = {{\sigma_{0|l}}^2}/{f_{0|l}}$ and $\alpha = {f_{0|l}}/{\theta}$. Using simple curve fitting over the plot of $f_{0|l}$ vs.\ $\sigma_{0|l}$, not shown in the paper due to space limitations, obtained from simulations with various $N$ and $l$, it can be seen that $\sigma_{0|l}  \approx f_{0|l} e^{-2f_{0|l}}$. The probability $\Pr\left(F_{0|l} < \epsilon\right)$ can then be approximated according to
\begin{eqnarray}\label{F-epsilon}
\Pr\left(F_{0|l} < \epsilon\right) \approx \frac{1}{C}\int_0^\epsilon \Gamma(x; e^{4f_{0|l}}, f_{0|l}e^{-4f_{0|l}}) dx. 
\end{eqnarray}

\subsection{Scenario 4: Small Cells Only, With Always-{\it on} $M$ SBSs}

\begin{figure}
\begin{center}
\includegraphics[scale = 0.4, trim= 60 110 130 180, clip]{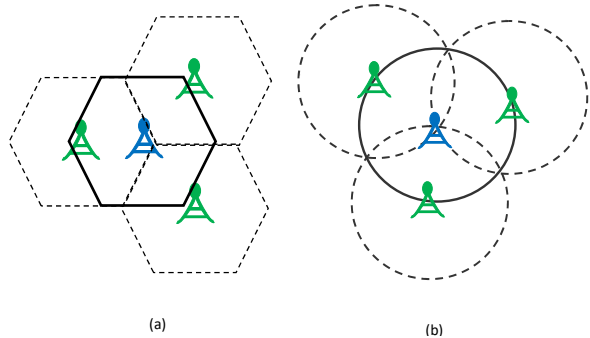}
\caption{Target cell has an overlay of $M$ always-{\it on} SBSs to provide coverage to the area. Randomly deployed SBSs are not shown. (a) small cells with same hexagonal area, (b) small cells with same circular area.}
\label{Fix3}
\vspace{-10mm}
\end{center}
\end{figure}

In this scenario, $M$ SBSs are strategically placed to cover the whole area as shown in Fig.\ \ref{Fix3} and other SBSs, not shown in the figure, are located randomly or according to the data demand. The fixed cells can also carry data. The users can be assigned to one of the reachable and active cells, among the always-{\it on} cells or the active cells among randomly deployed $N-M$ cells, with equal probability.

\begin{theorem}\label{beta-ns3-th}
The activation probability $\beta_4$ of the target cell, also for each of the $N-M$ randomly deployed cells, for scenario 4 is given as \vspace{-2mm}
\begin{eqnarray}
\beta_4= \sum_{l = 0}^{N-M} \beta_4^l {(1-\beta_4)}^{(N-M-l)} 
\left[1- p_{0_4}\left(P_{ss-on}(M)\right)^M \left(P_{ss_2}(M)\right)^l\right],\label{beta-ns3}
\end{eqnarray} 
where $p_{0_4} = \Pr(\text{no traffic than that served by M SBSs in }f_{0|l}A)$, $P_{ss-on}(M)$ is given in Corollary \ref{P-sat-on}, and $P_{ss_2}(M)$ is given in Theorem \ref{P-sat2}.
\end{theorem}
\begin{proof}
Consider the scenario with $l$, $0 \leq l \leq N$, active cells and $M$ always-{\it on} cells. The target cell will not be turned {\it on} if and only if the traffic from users in $f_{0|l}A$ remains within the serving capacity of $M$ always-{\it on} cells, and $M$ always-{\it on} SBSs and $l$ active SBSs remain unsaturated. Similar to the proof of Theorem \ref{beta-nsm-th}, the probability of the former event is given by
\[p_{0_4} = \sum_{j=0}^\infty\frac {(\lambda f_{0|l}A)^j} {j!} e^{-\lambda f_{0|l}A}\Pr\left( \sum_{i=0}^jY_i \leq N_{0_4} \right) ,\] 
where $N_{0_4}$ is the maximum RBs allocated by $M$ always-{\it on} SBSs to the traffic generated in $f_{0|l}A$. 
The probability that $M$ SBSs remain unsaturated is $\left(P_{ss-on}(M)\right)^M$, and that for $l$ active SBSs is $\left(P_{ss_2}(M)\right)^l$. Using symmetry in the problem and summing up all possible cases, we arrive at \eqref{beta-ns3}, which concludes the proof.  
\end{proof}
Similar to Theorem \ref{beta-nsm-th}, accurate characterisation of $N_{0_4}$ is not available. We propose an approximation using the relative saturation capacity idea presented in Section \ref{the-model}, i.e., $N_{0_4} \approx \lfloor\eta N_{\rm small}f_{0|l}\rceil$, where $\eta \geq 1$ compensates for the fact that some area could be served by multiple always-{\it on} SBSs. For example, $\eta = 1.17$ for the circular cell case and $\eta = 1$ for hexagonal cells.

\begin{corollary}
If small cells have hexagonal areas as shown in Fig.\ \ref{Fix3}-(a), the activation probability $\beta_4$ of the target cell, also for each of the $N-3$ randomly deployed cells, is given as \vspace{-2mm}
\begin{eqnarray}
\beta_4 = \sum_{l = 0}^{N-3} \beta_4^l {(1-\beta_4)}^{(N-3-l)} \left[1\hspace{-1mm}- p_{0_4}\hspace{-1mm}\left(P_{ss-on}(3)\right)^3 \left(P_{ss_2}(3)\right)^{l}\right]\hspace{-1mm},\label{beta-ns3-hex}
\end{eqnarray} 
where $p_{0_4}$ is given in Theorem \ref{beta-ns3-th}, $P_{ss-on}(3)$ is given in Corollary \ref{P-sat-on} with $A_0 = 1$, and $P_{ss_2}(3)$ is given in Theorem \ref{P-sat2} with $A_1 = 1$.
\end{corollary}
\begin{proof}
The proof is similar to the proof given for Theorem \ref{beta-ns3-th} with $M=3$. In this case, there are $3$ always-{\it on} cells which do not overlap each other, and hence $A_0 = 1$ (i.e., see Fig.\ \ref{Fix3}-(a)). Users in the target cell can be assigned to $l$ active SBSs or only one of the $3$ always-{\it on} SBSs, and therefore $A_1 = 1$ for $P_{ss_2}(3)$ in \eqref{beta-ns3-hex}. Summing up over all possible cases, we arrive at \eqref{beta-ns3-hex}.
\end{proof}

\begin{corollary}\label{beta-ns3-circular}
If small cells have circular areas as shown in Fig.\ \ref{Fix3}-(b), the activation probability $\beta_4$ of the target cell, also for each of the $N-3$ randomly deployed cells, is given as \vspace{-2mm}
\begin{eqnarray}
\hspace{-5mm}\beta_4= \sum_{l = 0}^{N-3} \beta_4^l {(1-\beta_4)}^{(N-3-l)} \left[1\hspace{-1mm}- p_{0_4}\left(P_{ss-on}(3)\right)^3 \left(P_{ss_2}(3)\right)^l\right]\hspace{-1mm},\label{beta-ns3-cir}
\end{eqnarray} 
where $p_{0_4}$ is given in Theorem \ref{beta-ns3-th}, $P_{ss-on}(3)$ is given in Corollary \ref{P-sat-on} with $A_0 = 0.66$, $A_1 = 0.34$, and $P_{ss_2}(3)$ is given in Theorem \ref{P-sat2} with $A_1 = 0.83$, $A_2 = 0.17$.
\end{corollary}

\begin{proof}
The proof follows from the similar lines as in the proof of Corollary \ref{beta-ns3-circular} by using standard equations for calculating overlapping areas for circular cells (i.e., see Fig.\ \ref{Fix3}-(b)).
\end{proof}
\vspace{-2mm}

\section{Numerical Analysis}\label{results}

In this section, we present the numerical results regarding power consumption under different scenarios described above. Unless stated otherwise, we have fixed the parameters as $N = 10$, $\bar{N_c} = 10$, $A = 3\times10^4\mbox{m}^2$ \cite{Marcano2018,Afshang2018}, $a = 1.8$, $b = 1$ms for the self-similar traffic\cite{Xiang2013}, which corresponds to $E[Y_i] = a\log(\tau/b) = 4$ for $\tau=10$ms. According to \cite{Marcano2018}, $20$ users can be assigned to a SBS and $70$ users to a macro BS in a typical network. We consider then $N_{\rm small} = 20E[Y_i]$ RBs and $N_{\rm macro} = 70E[Y_i]$. $A_c$, the area of each cluster is taken as $4A$ to allow all $N+1$ SBSs to lie within the cluster. Power consumption model parameters are selected as $P_{\rm macro} = 1000$W, $\rho_1 = 6$, $\rho_2 = 18$, $\rho_3 = 100$, $\rho_4 = 1.15$ \cite{Zaidi2015,Auer2011,Alhumaima2017}.  These values are close to the current state of the art of small cell deployments. $\epsilon$ is selected as $1\%$. We assume that all BSs have circular cell areas only for model validation and compare the corresponding theoretical results with the simulated probabilities. For the rest of the analysis, we only assume that all small cells have similar area $A$. The exception is scenario 4, where we assume 3 fixed always-{\it on} small cells to cover the area and calculate the activation probability using Corollary \ref{beta-ns3-circular}.

\subsection{Model Validation}\label{section_valid}

\begin{figure}
\begin{center}
\includegraphics[scale = 0.65, clip = true]{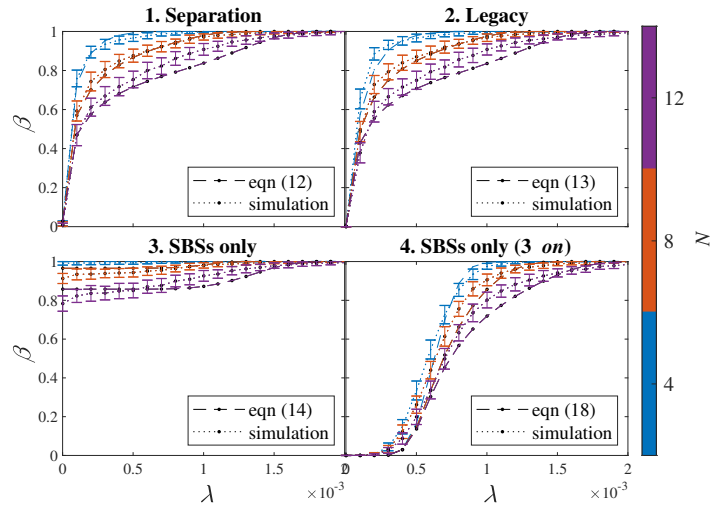}
\caption{Activation Probability $\beta$ versus $\lambda$ (user density) for scenarios 1, 2, 3, and 4. Here $\beta$ can be $\beta_1$ (separation), $\beta_2$ (legacy), $\beta_3$ (small cells only 1) and $\beta_4$ (small cells only with 3 {\it on} SBSs) relative to the scenario.}
\label{beta-123}
\vspace{-10mm}
\end{center}
\end{figure}

In order to verify Theorems \ref{beta-s-th}, \ref{beta-nsm-th}, \ref{beta-ns-th} and \ref{beta-ns3-th}, we simulated the scenarios in MATLAB. All SBSs are assumed to have circular coverage with radius $R = 100$m. The target SBS is assumed to be located at the origin  and $A_c$, the area of a cluster, is selected as $\pi(2R)^2$. Each simulation run generates one instance of all scenarios, i.e., location of $N$ SBSs, users and respective traffic, for the parameter settings, assigns users randomly to reachable BSs, activates enough SBSs in scenario 3 to achieve sufficient coverage, and saves the status of the target SBS once all users are assigned. The activation probabilities are calculated by summing target SBS' statuses for all scenarios and normalizing over total $T_s$ simulation runs. 

Figure \ref{beta-123} shows the activation probabilities calculated via theoretical expressions and relevant simulations for $N = 4, 8, 12$ and $\lambda = 0.1^{-5}, 0.201^{-3}, \cdots, 0.002$ with self-similar user traffic. The $y$-axis in Fig.\ \ref{beta-123} is $\beta$, representing $\beta_1$, $\beta_2$, $\beta_3$, or $\beta_4$ corresponding to the scenarios of separation architecture, legacy architecture, SBSs only network, or SBS only network including 3 {\it on} cells for coverage, respectively. Each curve represents a specific value of $N$. Plot colors  represent $N$ as shown by the color bar in Fig.\ \ref{beta-123}. Activation probabilities are calculated from simulations with $T_s = 100$ runs and are averaged over repeated $100$ trials. In particular, each trial consisted of $100$ simulation runs to generate a sample activation probability curve, and these curves are averaged over $100$ trials to obtain error bars for each data point. The error bars in Fig.\  \ref{beta-123} represent a single standard deviation on both sides of the mean value of $\beta$ from simulations. The compact error bars in Fig. \ref{beta-123} show that the simulated values of activation probabilities lie within a small range and closely follow the respective theoretical values.

\subsection{Energy Consumption Analysis}
\label{daily-profile}

An average data traffic profile over a day, denoted by $\Lambda$, is presented in \cite{ITU2015}. The report \cite{ITU2015} also shows the breakdown of different types of traffic, such as computing, streaming, etc. Most of the traffic types show higher usage during the day and very less during nights. In general, the time of the day affects the number of the active users in the system and the proportional use of the specific traffic type remains stable throughout the day. We selected to vary $\lambda$, the density of the users, with $\Lambda$ and kept the traffic model parameters $a$, $b$, and $\mu$ constant (cf.\ Section IV-C). Smaller $\Lambda$ results in fewer users in the system, which further results in less load. 

\begin{figure}
     \centering
     \begin{minipage}{.48\textwidth}
         \centering
         \includegraphics[width=\textwidth,height=5cm]{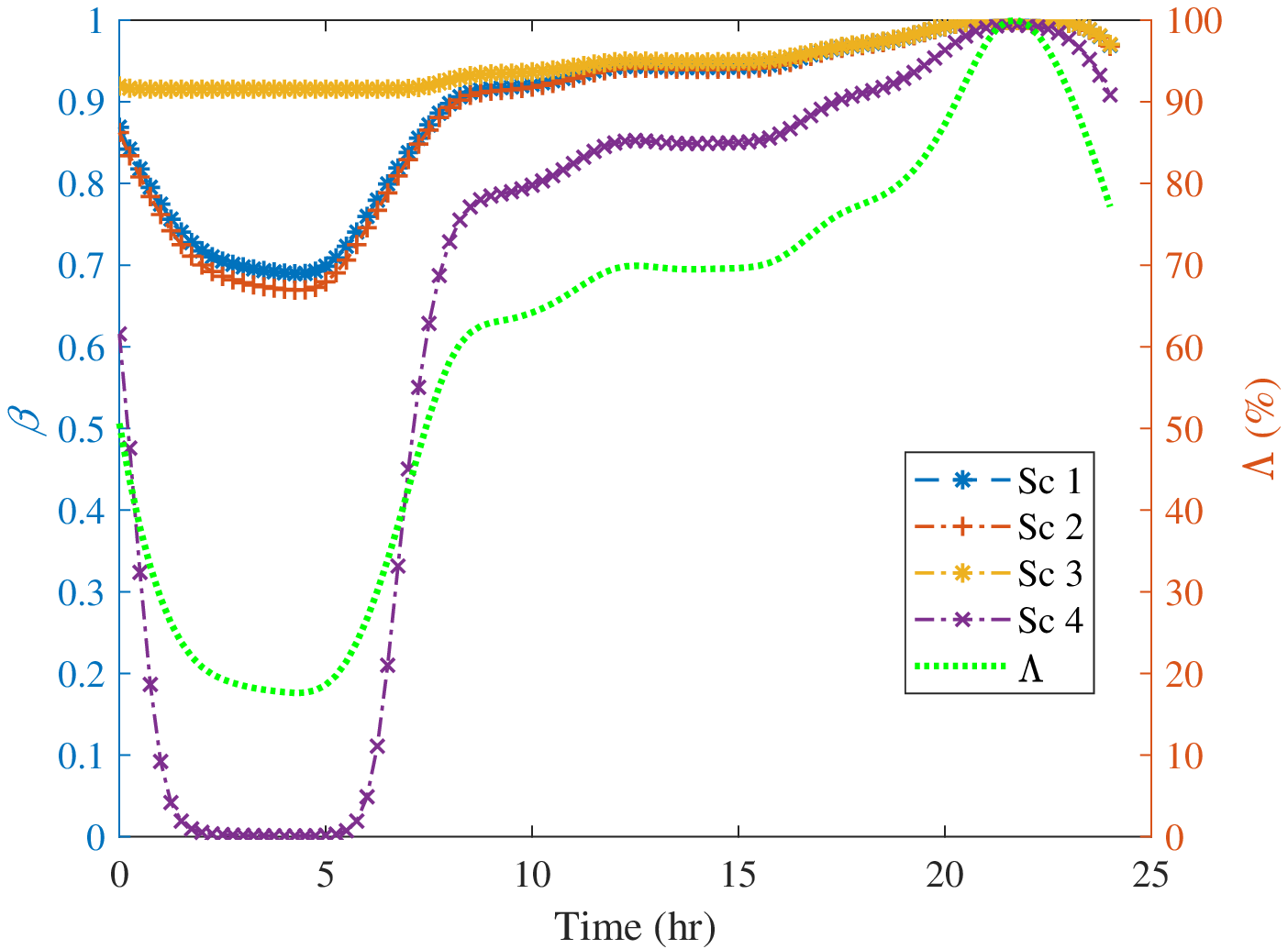}
         \captionof{figure}{Activation probabilities for scenario 1, 2, 3, and 4 and the normalized data traffic profile for $24$ hours. $\beta_1$ and $\beta_2$ mostly overlap.}
         \label{load-activation}
     \end{minipage}
     \vspace{-4mm}
     \hfill
     \begin{minipage}{.48\textwidth}
         \centering
         \includegraphics[width=\textwidth,height=5cm]{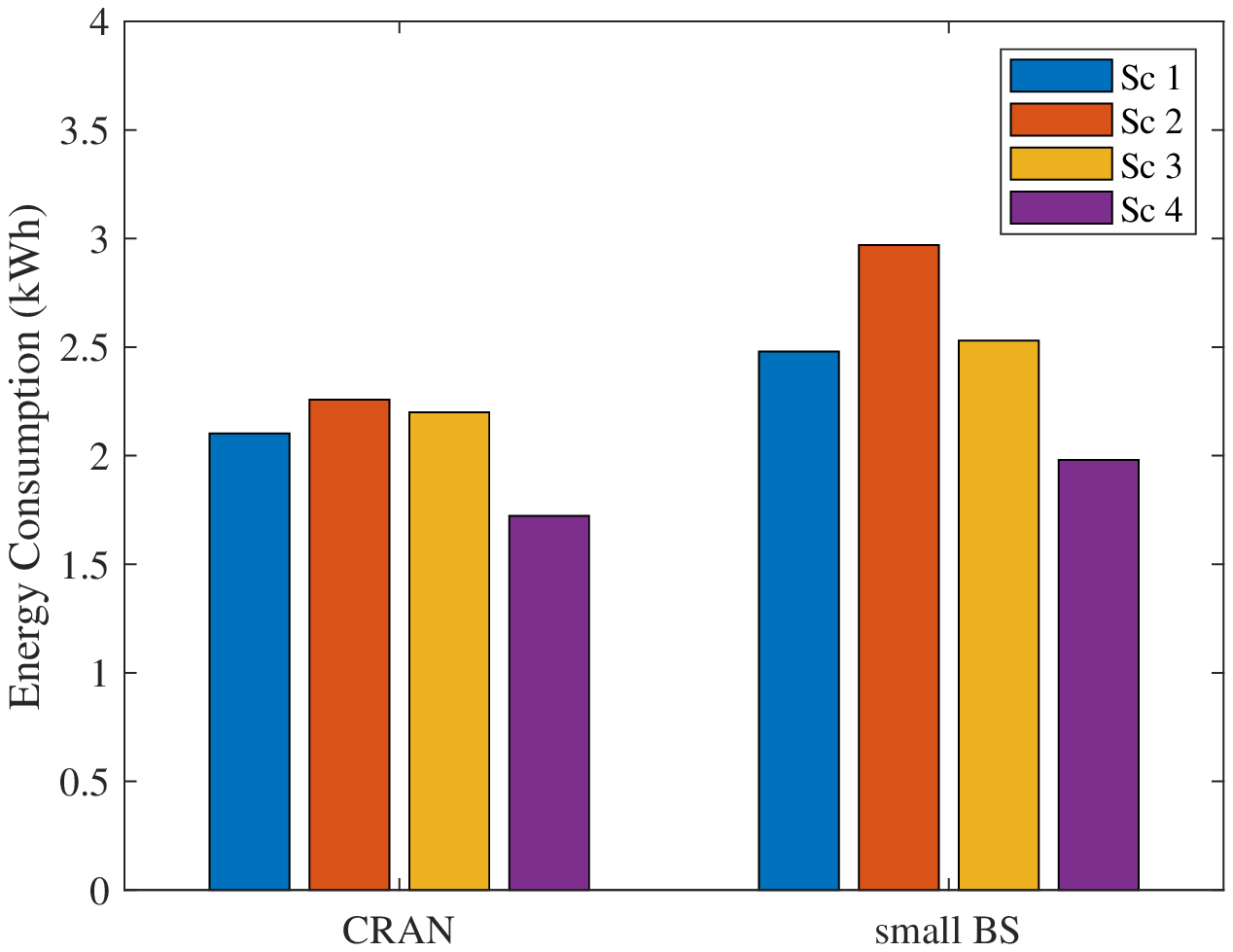}
         \captionof{figure}{Daily energy consumption for all four scenarios calculated from activation probabilities.}
         \label{EC}
     \end{minipage}
        \vspace{-4mm}
\end{figure}

We select $\lambda_{max} = 0.0016$ as the minimum value of $\lambda$ which yields $\beta_i = 0.99 \approx 1, \forall i$, interpreted as the full load. The user density for our experiments will be $\lambda = \lambda_{max}\Lambda/100$. The normalized traffic profile $\Lambda$ and associated $\beta_1$, $\beta_2$, $\beta_3$ and $\beta_4$ are given in Fig.\ \ref{load-activation}. 


In this figure, $\beta_3 > \beta_1$ for low traffic because the SBSs in scenario 3 are also responsible for providing coverage. However, $\beta_1 \approx 0.69$, which can be interpreted as around $70\%$ active cells. This is the major cause of separation architecture not being as beneficial as expected. Although, the traffic demand is low ($\approx 20\%$), but the associated users are randomly distributed within the area and multiple small cells are required to be active. 

The signaling BS in scenario 1 can also be designed to provide some data transmissions as well so all the data BSs can be turned {\it off} during low-utilization phase. Scenario 4 can be a relevant example where we have 3 SBSs to cover the area. $\beta_1$ is not expected to be lower than $\beta_2$ and $\beta_4$ as macro cell of scenario 2 and always-{\it on} cells can serve some load. Scenario 1 can only be power efficient than scenario 2 and 4 if the signaling only macro BS consumes much less power than the standard macro BS and always-{\it on} SBSs in comparable area. 

Using the activation probabilities over 24 hours and the power consumption model given in Section \ref{PC-levels}, we add the energy consumption in kWh for all cells in scenarios 1-4. As an example, the total energy consumption in a day for scenario 1 is 
\begin{eqnarray}\label{EC-scenario1}
EC_1 = \frac{(N+1)}{\phi}\sum_{i=1}^{24\phi}(\beta_{1_i}P_{\rm small}+ (1 - \beta_{1_i})P_{\rm sleep}) + 24 A\frac{P_{\rm sig-macro}} {\bar{N_c}A_c},
\end{eqnarray} 
where $\beta_{1_i}$ is the activation probability for scenario 1 at the $i$th sample instant, $\phi$ is the number of samples in an hour and $\bar{N_c}$ is the mean number of clusters, each with area $A_c$, in a macro cell. The energy consumption of other scenarios and also for CRAN scenarios are also calculated in similar manner. Energy consumption values for scenarios 1-4 are given in the bar chart of Fig.\ \ref{EC}. As expected, CRAN based scenarios have better energy consumption than the scenarios with SBSs. CRAN may be more energy efficient in practice as it offers better energy saving opportunities through BBU sharing and more sophisticated alternatives of shutting down equipment.

The most important result of Fig.\ \ref{EC} is the marginal energy savings in scenario 1, i.e., the separation architecture, when compared against scenarios 2 and 3. Separation architecture saved $489$Wh over a day compared to the legacy architecture which turned out to be $16.48\%$ savings. This saving can be considered reasonable but when compared for the CRAN case, the separation architecture saves a mere $6.85\%$ against the legacy architecture. Comparing with scenario 3, comprised of small cells only, separation architecture saves $49.5$Wh in a day which is approximately $1.96\%$ savings. For the CRAN case, the energy saving of separation architecture stands at $4.41\%$ against scenario 3. On the other hand, scenario 4, with SBSs and CRAN, provides $33.31\%$ and $23.68\%$ respective savings compared to scenario 2 and $21.7\%$ savings compared to scenario 3. In scenario 4, we assumed 3 always-{\it on} small cells to provide coverage to the analysis area. In practical situations, we may need more than $3$ SBSs due to uneven terrain.


\subsection{Sensitivity to the parameters} 

In order to be confident about our conclusions, we performed comprehensive sensitivity analysis by considering the effects of almost all parameters in our study. We use {\it \% Energy Savings} as a figure of merit. The \% Energy Savings, $\gamma_{ij}$, for scenario $i$ with respect to scenario $j$ is defined as \[\gamma_{ij} = \frac{EC_j - EC_i}{EC_j}\times 100,\] where $EC_i$ is the energy consumption of scenario $i$ computed over a day and $i = 1$ and $j = 2-4$.

\subsubsection{Traffic Model}

Table \ref{traffic_esavings} shows $\gamma_{ij}$ when traffic was generated using Poisson as well as self-similar distributions. We have varied the model parameters in both models. Set of parameters with same ID across the models have same mean traffic $E[Y_i]$. $E[Y_i]$ is $\mu\tau$ for Poisson model. We can see that the \% energy savings remain small across the varied traffic for separation architecture and there are no significant variations specially in $\gamma_{12}$ for both cases of SBSs and CRAN. Moreover, Scenario 4 remains superior to the separation architecture as shown by the negative $\gamma_{14}$. We expected this result as variations in user traffic effect all scenarios in almost a similar manner and thus the relative measure of \% energy savings remains almost the same. 
\begin{table}{
  \begin{center}
    \caption{Traffic Models and $\gamma_{ij}$ (SBS, CRAN).}\vspace{-2mm}
    \label{traffic_esavings}
    \begin{tabular}{c|l|c|c|c}
	\hline \hline
      \multicolumn{2}{c|}{Traffic Model} & $\gamma_{12}$ (\%) & $\gamma_{13}$ (\%)& $\gamma_{14}$ (\%)\\
	\hline
     \multirow{3}{1cm}{Self-similar} & $1- (a, b) = (1.8, 1\mbox{ms})$ & 16.48, 6.85 & 1.96, 4.41 & -25.24, -22.05 \\
     &$2-(a, b)=(1.8,0.1\mbox{ms})$& 16.81, 7.28 & 1.78, 4.24 & -25.56, -22.36 \\
     &$3-(a, b) = (2, 1\mbox{ms})$ & 16.53, 6.9 & 1.89, 4.35 & -25.66, -22.46 \\\hline
     \multirow{3}{*}{Poisson} & $1- \mu = 400$ & 16.48, 6.85 & 1.96, 4.41 & -25.24, -22.05\\
     & $2- \mu = 828$ & 16.81, 7.28 & 1.78, 4.24 & -25.56, -22.36\\
     & $3- \mu = 439$ & 16.53, 6.91 & 1.89, 4.35  & -25.66, -22.46\\
      \hline\hline
    \end{tabular}
  \end{center}}
  \vspace{-8mm}
\end{table}  

\subsubsection{Hard Core Point Process (HCP)}\label{results_HCP}

In a HCP, the SBSs are at least $r_d$ apart from each other \cite{Elsawy2013}. This distribution is also called Mat\'ern I HCP and is based on an original PPP with minimum distance constraint. We calculate activation probabilities for each scenario through simulation experiments.

Table \ref{HCP_esavings} shows $\gamma_{ij}$ when SBSs in all scenarios follow HCP with specific $r_d$. Traffic is considered to be self-similar with $(a, b) = (1.8, 1\mbox{ms})$. The \% energy savings of separation architecture with respect to the other architectures is not significantly different from that computed with PPP spatial distribution (row 1 of Table \ref{traffic_esavings}). Separation architecture is still marginally better than scenarios 2 and 3, and significantly worse than scenario 4.

As $r_d$ increases, we observe the decrease in activation probabilities as the SBSs are more spread out and cover a larger area together, but it happens for all deployment scenarios. The \% energy savings with respect to legacy scenario remain almost the same for both SBS and CRAN cases. $\gamma_{13}$ decreases indicating that scenario 3 is able to close the gap as it now requires smaller number of SBSs to provide coverage. The decrease is more pronounced in case of SBSs where the power consumption per unit of equipment is higher than the case of CRAN. Scenario 4 remains superior for all values of $r_d$.

\begin{table}{
  \begin{center}
    \caption{HCP and $\gamma_{ij}$ (SBS, CRAN).}\vspace{-2mm}
    \label{HCP_esavings}
    \begin{tabular}{c|c|c|c}
	\hline \hline
      $r_d$ (m) & $\gamma_{12}$ (\%) & $\gamma_{13}$ (\%)& $\gamma_{14}$ (\%)\\
	\hline
     20& 16.36, 6.87& 0.29, 2.75& -22.64, -19.57 \\
     40& 16.45, 6.9 & 0.27, 2.74& -22.57, -19.49\\
     60& 16.33, 6.78 & 0.002, 2.48& -23.37, -20.27\\
     \hline \hline
    \end{tabular}
  \end{center}}
  \vspace{-5mm}
\end{table}  

\subsubsection{System Model Parameters}

\begin{figure}
\begin{center}
\includegraphics[width = 8.5cm,height = 7cm, clip = true]{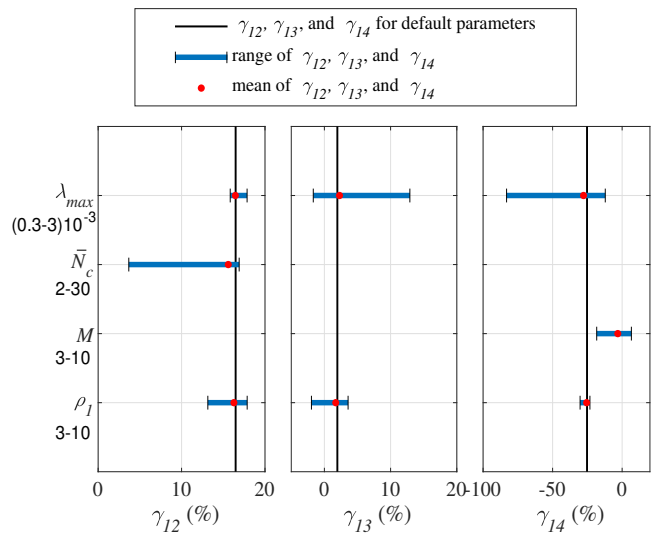}
\caption{The variation in $\gamma_{12}$, $\gamma_{13}$, and $\gamma_{14}$ when $\lambda_{max}$, $\bar{N_c}$, $M$, and $\rho_1$ are varied between the given ranges. All the parameters are kept at default values while one is varied.}
\label{gamma-lambda}
\vspace{-10mm}
\end{center}
\end{figure}

In Fig.\ \ref{gamma-lambda}, we varied $\lambda_{max}$ from $0.0003$ to $0.003$, $\bar{N_c}$ from $2$ to $30$, $M$, the number of fixed always-{\it on} SBSs in scenario 4, from $3$ to $10$, and $\rho_1$, the ratio between power consumption of a normal macro BS and a signaling only macro BS, from $3$ to $10$. In each experiment, other parameters are fixed at default values while one is varied.  Moreover, since relative merit of alternative architectures remains similar across SBS and CRAN, we are only focusing on networks with SBSs instead of RRH to study the effects of parameter values. CRAN also produced similar conclusions. 

The thick black lines show the respective energy savings for default parameter values, i.e., $M = 3$, $\lambda_{max} = 0.0016$, $\bar{N_c} = 10$, and $\rho_1 = 6$. We observed significant variations in \% energy savings across the scenarios, although, the variations are not always linear with the varied parameter. The mean values of $\gamma_{ij}$, shown with the red circle, are indicative of this fact. 

The key takeaway point of Fig.\ \ref{gamma-lambda} is that the maximum observed \% energy savings of separation architecture stays under $20\%$ in all cases. Variations in $\lambda_{max}$ caused $\gamma_{12}$ and $\gamma_{13}$ to improve but only for very small values of $\lambda_{max}$ where maximum resource utilization for scenario 1 falls to $60\%$ at peak load in a day and almost $40\%$ of the equipment is never used. On the other hand, $85\%$ of the small cells in scenario 3 are active to provide coverage. Scenario 4 remains superior for all values of $\lambda_{max}$. As shown in Fig.\ \ref{gamma-lambda}, variations in ${\bar N_c}$ and $\rho_1$ only caused marginal improvements in \% energy savings, if any. Effects of varying $N_{\rm macro}$ can also be explained by the variations in $N_c$. 

Figure \ref{gamma-lambda} also shows the impact of increasing $M$ on $\gamma_{14}$. Here without considering any particular shape of the cell, we assume that $M$ cells, each with area $A$, are needed to cover the analysis area. For simplicity, we assume that the cells do not intersect each other. Overlapping cells, in general, allow more users to be served in the area lowering the activation probability and corresponding energy consumption of the system. Our assumption, though may not be true in practical scenarios, only makes the scenario 4 worse and not better. This reason causes a higher value of $\gamma_{14}$ shown in Fig.\ \ref{gamma-lambda} for $M=3$ than the default $\gamma_{14}$ shown by thick black line, and also reported by row 1 of Table \ref{traffic_esavings}. Figure \ref{gamma-lambda} shows that as $M$ increases, scenario 4 becomes slightly worse than the separation architecture. One important point here is that uneven terrain would also affect the signaling macro cell in a separation architecture and we may need multiple signaling cells to counter the signal attenuation through obstacles. We did not consider multiple signaling macro BSs in Fig.\ \ref{gamma-lambda} giving an advantage to the separation architecture.

Among the system model parameters, some impact all the scenarios in similar manner. For example, changing $N$, the number of overlapping small cells, $N_{\rm small}$, the saturation capacity of small cells, and $A$, the area of small cells, changes the capacity and load of the system. Which shifts $\beta$ vs.\ $\lambda$ (cf.\ Fig.\ \ref{beta-123}) curves to either left or right but the relative energy consumption of different scenarios would remain the same. Similarly, varying $\rho_3$, i.e., the ratio between power consumption of macro and SBS, and $\rho_4$, the ratio between power consumption of SBS and small RRH, has no effect over \% energy savings as variation in numerator is canceled out by that of the denominator.  

\subsubsection{Location Based Daily Traffic Profile}

We also tested the scenarios under study with different daily traffic profiles. A detailed study on various types of weekly data traffic profiles is performed in \cite{Xu2017} for different environments, such as residential, office, and transport areas. Transport area profile is particularly interesting because the traffic in the night is very close to zero. All data traffic profiles are normalized by the overall peak value and used in the calculation of activation probabilities and power consumption values for scenarios 1, 2, 3, and 4 in a similar manner and with similar parameters as in Section \ref{daily-profile}. $\lambda_{max}$ is chosen as $1.4\times10^{-5}$ for residential environment, $2.8\times10^{-5}$ for office area, and $5\times10^{-5}$ for transport area profile. These are the values of $\lambda$ which yield full utilization of resources. 

\begin{figure}
\begin{center}
\includegraphics[scale=0.7, clip = true]{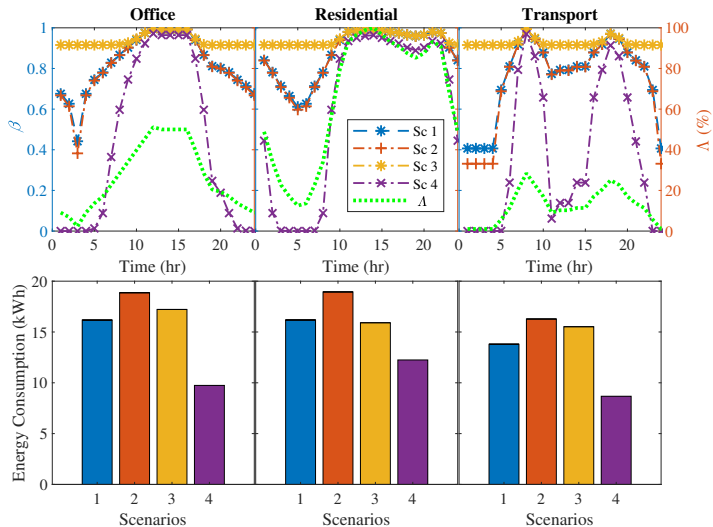}
\caption{Normalized data traffic profiles, activation probabilities, and weekly energy consumption for residential area, office area, and transport area \cite{Xu2017}.}
\label{alphas}
\vspace{-12mm}
\end{center}
\end{figure}

The traffic profiles for one day are shown in top panels of Fig.\ \ref{alphas} along with the activation probabilities of all scenarios. Our analysis took weekly profile \cite{Xu2017} into consideration. Due to space constraints the traffic over a whole week is not shown here. The weekly energy consumption in kWh is given in Fig.\ \ref{alphas}. In all of the environments, scenario 4, the small cells only network with some fixed always-{\it on} cells, has the lowest consumption. The energy consumption of scenario 1, the separation architecture, is only marginally better. 

\section{Related Work}\label{rel-work}

First presented by Greentouch consortium (www.greentouch.org) \cite{Rittenhouse2012}, the idea of separation architecture was studied under EU FP7 project EARTH \cite{Godor2012} and by Ericsson and NTT DOCOMO under the names `lean carrier' \cite{Hoymann2013} and `Phantom cell concept' \cite{Ishii2012} and followed up with evaluation studies \cite{Mukherjee2013}. The idea is also included in 3GPP release 12 under the concept of New Carrier Type \cite{Siedel2013}. The focus of this section is only on the energy/power consumption studies for separation architecture and not on the related development of the CDSA (Control-Data Separation Architecture) technology \cite{Pat2019,Taufique2019}. Most of the efforts to understand the margins of energy savings reported very attractive figures. We found some issues with the assumptions in all of the studies. Also, they all have compared against all-{\it on} scenario and did not use any energy management scheme with the legacy network.   

The analysis reported in \cite{Filippini2017} shows that depending on the daily load profile, the energy efficiency can be improved by more than $50$ times comparing to the legacy systems. We think that these results could be due to under-utilized separation architecture. The paper reported $20\%$ activation probability at full load which means that $80\%$ of the equipment was never used. When this activation probability is compared against $100\%$ {\it on} legacy systems, the discrepancy in energy efficiency is greatly inflated. Moreover, the data BSs in the separation architecture do not have finite capacity resulting in very low activation probability. The study in \cite{Godor2012} under EU FP7 project EARTH shows that the opportunity of improved sleep cycles in the separation architecture have saving potential of $85-90\%$ when compared to the legacy systems. This study employs completely silent sub-frames, i.e., $0\%$ load, to measure the saving potential of separation architecture as compared to the always-{\it on} legacy BS.

In a follow-up paper on Phantom cell concept, the authors calculated spectral efficiency and energy efficiency of Phantom cells (on-demand data cells in a separation architecture) and compared it to the scenario with small cells \cite{Mukherjee2013}. Both scenarios assumed macro cell overlay. In small cells, the same channels are used for macro and SBSs, whereas Phantom cells used different channels than macro cell, which improved its spectral efficiency. Improved spectral efficiency also translated into better energy efficiency, defined as bit/J. The power consumption comparison in \cite{Wang2014} of the separation architecture is done with that of a single macro cell scenario and it showed that $50\%$ or more reduction is possible. We remark that the savings are probably the result of using small cells in separation architecture than a macro cell. The relative energy savings, of multiple small cells when compared against macro cell providing similar performance, depends on many factors and one is not always better than another.  

The power consumption of control-data separated heterogeneous CRAN (H-CRAN) is compared against a conventional HetNet, instead of a non-separated H-CRAN, in \cite{Liu2016} and the simulation results show $16\%$ improved power consumption. Interestingly, the paper did not include {\it sleep} or {\it off} modes with the data BSs in their study and the improvement in the energy efficiency is mainly due to the sharing of centralized BBU pool in H-CRAN, centralization of control signaling, and also efficient resource allocation with reduction in control overhead.  

The feasibility study reported in \cite{Ternon2013} also did not consider the finite capacity of BSs rather they showed, through simulations, that low number of users would trigger a low activation probability for on-demand BSs. The activation probability is not compared against that of the current systems. 
In a recent study \cite{Kang2017}, the authors developed energy management strategies over the separation architecture using {\it sleep} and {\it off} states of BSs and compared them with each other. There was no comparison with the energy consumption of legacy system. Similarly, the recent paper \cite{Liang2019} calculates and optimizes coverage probability of signaling BSs under non-line-of sight (NLoS) and energy efficiency of data BSs under LoS and NLoS conditions but did not compare the energy efficiency with that of a legacy system.


\section{Conclusions}\label{conc}
In this paper, we have modeled energy consumption for next generation HetNets with and without logical separation of control and data transmissions. We expected to observe very low energy consumption with separation architecture, but the results indicated otherwise and the maximum \% saving was under $18\%$ when compared with a legacy network. For a CRAN based setting, this margin was further reduced to under $7\%$. When we fixed minimal number of SBSs to cover the area in a legacy architecture, with small cells only and without the separation, and kept them always {\it on}, the non-separation architecture outperformed the separation architecture by huge margin. One may argue that in practice, we may need more small cells to cover the area because of uneven terrain and signal blockages due to buildings. However, this would affect the signaling macro cells in a separation architecture as well. 

The major reason for the discrepancy in our expectations and results is the non-negligible requirement of active cells during low utilization period of the day. Our results show that around $70\%$ of the SBSs in a separation architecture are needed to be {\it on} during nighttime according to the daily data traffic profile. For this type of traffic, a CRAN with shared BBU pool is more efficient. Also, the signaling infrastructure should also be designed to carry data during low-traffic phase. As future extension of the current research, we are planning to develop a framework for detailed dimensioning of a signaling BS. This framework will be based on generalized energy consumption model taking the transmission rates, channel state, and multiple access interference into account. The model will be able to optimize signaling BS' data capacity for given wireless network and operating conditions with respect to energy efficiency.

As another future extension of the current paper, we are also developing mathematical expressions for activation probabilities of target SBS when the overlapping SBSs are deployed according to a Hard Core Point Process. As overlapping small cells increase, the problem becomes very complex due to possible combinatorial nature of overlapping coverage and loss of independence in deployment of SBSs.
\vspace{-5mm}
\section*{Acknowledgement}

We are grateful to the editor's and anonymous reviewers' valuable feedback. It helped to improve the quality and presentation of our study. 
\vspace{-5mm}
\bibliographystyle{IEEEtran}

\bibliography{SplitRAN_Ref}

\end{document}